# Quantum state transfer in frequency down-conversion via nondegenerate optical parametric amplifier

Yunfei Bai,[1] Junxiang Zhang,[1,*] and Shi-Yao Zhu[1,2,3]

[1]*The State Key Laboratory of Quantum Optics and Quantum Optics Devices, Institute of Opto-Electronics, Shanxi University, Taiyuan 030006, China*

[2]*Beijing Computational Science Research Center, Beijing 100084, China*

[3]*Department of Physics, Hong Kong Baptist University Hong Kong, China*

[*]*junxiang@sxu.edu.cn*

**Abstract:** By utilizing the nondegenerate optical parametric amplifier, the quantum state transfer from a pump state with high frequency to a signal state of lower frequency is studied theoretically. The noiseless state transfer is characterized by frequency conversion efficiency and noise figure. It is shown that the quantum state can be well preserved during the frequency conversion as long as the optical parametric amplifier is operated far below the threshold, and meanwhile injected with an input idler field with a power density about ten times greater than that of the pump. The dependence of the noise figure and conversion efficiency on the idler amplitude, the analyzing frequency and the cavity extra loss is also discussed.

## 1. Introduction

Parametric frequency conversion is an attractive technique to generate coherent tunable radiation in the optical range, to transform radiation from one frequency to others, and to generate nonclassical states [1,2], it has become a major ingredient in quantum network for large-scale quantum information process [5]. The quantum network, which consists of quantum nodes and channels, needs the atoms to be used as quantum nodes to process and store quantum states locally [6] and the photons acted as quantum channels to link the separated nodes for the exchange of quantum information [7]. Fundamentally, this endeavor is the quantum interface that converts quantum states from one physical system to those of another in a reversible fashion. Such quantum connectivity can be achieved by optical interaction of photons and atoms [8]. Therefore, the quantum state transfer between atoms and photons was theoretically and experimentally developed in cavity QED [9,10], electromagnetically induced transparency (EIT) [11], Raman [12] and four-wave mixing processes [13].

It is well known that the transmission wavelengths for photons in telecommunication optical fibers are 1310 nm and 1550 nm, and the atoms absorb/emit photons at a different wavelength, e.g. 800 nm for alkaline atoms [14], thus in order to build a quantum interface, an optical frequency conversion is needed to couple the photons of communication band with atoms. Of course, for faithful quantum frequency conversion, a noiseless quantum state would be transferred during this procedure [15].

The information-preserving unitary transformation between two different frequencies can be realized via particle annihilation or creation process in a nonlinear frequency conversion, the optimum candidate is frequency up-conversion because of noise-free and 100% conversion efficiency [16]. Since the conception for noise-free photon frequency up-conversion was proposed and experimentally realized [15, 17], it has been extensively developed for high efficiency single photon detection of infrared (IR) photons [18], and the single photon detection efficiency is further enhanced via frequency upconversion in a cavity or in a waveguide[19, 20]. Very recently, the frequency up-conversions at the single photon level via four-wave mixing in fibers [21, 22] showed that the frequency up-conversion becomes an essential technology for quantum network. However, the noiseless frequency conversion of a quantum state is difficult to be realized in the down-conversion process, because the process is normally considered as an amplification process with unavoidable amplified quantum noise (spontaneous emission) [23,24]. Recently, the proof-of-principle experiment on spontaneous frequency down-conversion of a traveling-wave light was demonstrated, and 1% conversion efficiency from a weak pump field to a signal field was obtained [25]. And very recently, the coherence-preserving photon frequency down-conversion has also been realized in waveguide material [26]. In addition to the case that the frequency down-conversion was realized in free space or waveguide, the frequency down-conversion in an optical parametric amplifier/oscillator (OPA/O) has been demonstrated as an efficient way to enhance the effective conversion for nonclassical state generation [27-29]. Motivated by this, in this paper, we present the frequency down-conversion of a quantum state with a nondegenerate optical parametric amplifier (NOPA). The quantum property of frequency down-conversion of a quantum state is characterized and quantified by the noise figure and frequency conversion efficiency. The condition for noise-free conversion is analyzed, and the dependence of conversion efficiency and signal conversion efficiency on the pump parameter, the amplitude of the injection of an idler mode and intracavity loss is also discussed. Based on the unique advantages of OPA/OPO in comparison to the spontaneous process, such as high nonlinear gain, low threshold for pump power, wide emission range and high stability, this scheme for quantum state transfer with OPA might be of great importance in the quantum information processing.

**2. Theoretical model**

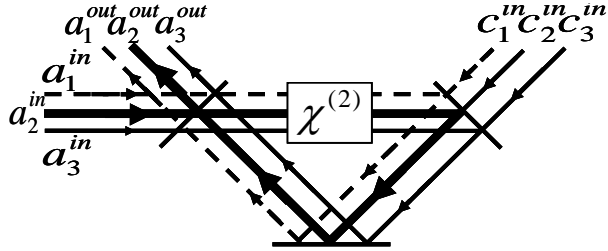

FIG. 1. The sketch of intracavity OPA process

The scheme of the frequency down-conversion of a quantum state utilizing the intracavity NOPA process is shown in Fig. 1. The system consists of three modes of the electromagnetic field coupled by a nonlinear crystal inside an optical cavity. There are three modes, signal $(\hat{a}_1)$, idler $(\hat{a}_2)$ and pump $(\hat{a}_3)$ with frequencies $\omega_1$, $\omega_2$ and $\omega_3$, respectively, where $\omega_3 = \omega_1 + \omega_2$ for energy conservation. If the input pump mode denoted by $a_3^{in}$ (a higher frequency quantum state) is converted to lower frequency signal mode $a_1^{out}$ (output signal mode) with 100% conversion efficiency and without any noise, then the requirements are

$$\delta^2 X(Y)_1^{out} = \delta^2 X(Y)_3^{in}, \tag{1}$$

$$\overline{I_1^{out}} = \overline{I_3^{in}}, \tag{2}$$

where $\delta^2 X(Y)$ is quadrature amplitude (phase) component fluctuation, and $\overline{I}$ is the average intensity of the corresponding modes.

For the quadrature components fluctuation and the average intensity of the quantum state (lower frequency) to be converted from a higher frequency, we use the noise figure (NF) $T_{X(Y)}$ and conversion efficiency $\eta$ to evaluate the performance of the frequency conversion, which are commonly defined as [13,30]

$$T_{X(Y)} = \frac{SNR[X(Y)_1^{out}]}{SNR[X(Y)_3^{in}]} = \frac{\left\langle X(Y)_1^{out} \right\rangle^2 / \left\langle \delta^2 X(Y)_1^{out} \right\rangle}{\left\langle X(Y)_3^{in} \right\rangle^2 / \left\langle \delta^2 X(Y)_3^{in} \right\rangle}, \tag{3}$$

$$\eta = \frac{\overline{I_1^{out}}}{\overline{I_3^{in}}}, \tag{4}$$

where $SNR[X(Y)^{in(out)}]$ is signal-to-noise ratio of the quadrature amplitude (phase) component of the input (output) mode. For an ideal frequency conversion of a quantum state, the signal-to-noise ratio on the output signal mode $a_1^{out}$ is identical to that of the input pump mode $a_3^{in}$, then we have $T_X = T_Y = 1$ and the average intensity of two modes should be equal, i.e. $\eta = 1$.

This scheme is similar to quantum nondemolition measurement (QND) and noiseless signal amplification, but there is a different between them. In the QND, the signal mode $S$ and the meter mode $M$ correspond here to $a_3^{in}$ and $a_2^{in}$, respectively. To make a measurement on $a_3^{in}$ without disturbing its evolution, the conditions, $\delta^2 a_3^{in} = \delta^2 a_3^{out}$ and $\delta^2 a_3^{in} = \delta^2 a_2^{out}$ are required, equivalently $1 < T_S + T_M < 2$ [31]. For the noiseless signal amplification, $T_S = 1$ is the only condition [32]. In our scheme, the input state's SNR ($a_3^{in}$) is transferred to the signal mode $a_1^{out}$ and the average intensity is also required to be equal.

Consider the NOPA process with perfect phase matching, zero detuning, and small losses, see Fig. 1, the evolution equations for this system can be given by [33]

$$\tau \dot{\hat{a}}_1(t) = -(\gamma_1 + \rho_1)\hat{a}_1(t) + \chi \hat{a}_2^+(t)\hat{a}_3(t) + \sqrt{2\gamma_1}\hat{a}_1^{in}(t)e^{i\theta_{10}} + \sqrt{2\rho_1}\hat{c}_1^{in}(t), \tag{5a}$$

$$\tau \dot{\hat{a}}_2(t) = -(\gamma_2 + \rho_2)\hat{a}_2(t) + \chi \hat{a}_1^+(t)\hat{a}_3(t) + \sqrt{2\gamma_2}\hat{a}_2^{in}(t)e^{i\theta_{20}} + \sqrt{2\rho_2}\hat{c}_2^{in}(t), \quad (5b)$$

$$\tau \dot{\hat{a}}_3(t) = -(\gamma_3 + \rho_3)\hat{a}_3(t) - \chi \hat{a}_1(t)\hat{a}_2(t) + \sqrt{2\gamma_3}\hat{a}_3^{in}(t)e^{i\theta_{30}} + \sqrt{2\rho_3}\hat{c}_3^{in}(t), \quad (5c)$$

where $\hat{a}_i^{in}$ $(i=1,2,3)$ denote the input amplitude operators, $\hat{c}_i^{in}(t)$ is the vacuum noise term corresponding to intracavity loss, and $\chi$ is the effective nonlinear coupling parameter and is proportional to the second-order susceptibility $\chi^{(2)}$ of crystal. The roundtrip time $\tau$ in the cavity is assumed to be same for all three fields. The total loss parameter is $\gamma_i + \rho_i$ $(i=1,2,3)$, where $\gamma_i$ are related to amplitude reflection and transmission coefficients of the coupling mirror, $\rho_i$ $(i=1,2,3)$ represent the extra intracavity loss parameters.

For simplicity, we assume the amplitude $\beta_3$ and $\beta_2$ for the input pump $\overline{\alpha_3^{in}}$ and injecting idler mode $\overline{\alpha_2^{in}}$ are real, the input of signal mode $\overline{\alpha_1^{in}}$ is vacuum, the initial phases for the three modes are $\theta_{10} = \theta_{20} = 0$, $\theta_{30} = \pi/4$, and the cavity transmission factor and the extra losses for the signal and idler modes are the same $\gamma_1 = \gamma_2 = \gamma$, $\rho_1 = \rho_2 = \rho$. Then, the steady-state equations of Eq. (5) are [34]

$$-(\gamma + \rho)\overline{\alpha_1}e^{i\theta_1} + \chi \overline{\alpha_2^*}\,\overline{\alpha_3}e^{i(\theta_3 - \theta_2)} = 0, \quad (6a)$$

$$-(\gamma + \rho)\overline{\alpha_2}e^{i\theta_2} + \chi \overline{\alpha_1^*}\,\overline{\alpha_3}e^{i(\theta_3 - \theta_1)} + \sqrt{2\gamma}\beta_2 = 0, \quad (6b)$$

$$-(\gamma_3 + \rho_3)\overline{\alpha_3}e^{i\theta_3} - \chi \overline{\alpha_1}\,\overline{\alpha_2}e^{i(\theta_1 + \theta_2)} + \sqrt{2\gamma_3}\beta_3 e^{i\pi/4} = 0, \quad (6c)$$

where $\overline{\alpha_1}, \overline{\alpha_2}, \overline{\alpha_3}$ are steady-state solution of intracavity modes $\hat{a}_1, \hat{a}_2, \hat{a}_3$ respectively. The above equations show that all the three solutions of $\overline{\alpha_1}, \overline{\alpha_2}, \overline{\alpha_3}$ are real numbers. The NOPA oscillation threshold $\varepsilon_{th}$ for pump mode is obtained when considering $\overline{\alpha_2^{in}} = 0$, and the required steady-state amplitude of the pumping mode and solutions are given by

$$\varepsilon_{th} = \frac{(\gamma_3 + \rho_3)(\gamma + \rho)}{\chi} \quad (7a)$$

$$\theta_1 = \theta_3 - \theta_2 = \theta_{30} - \theta_{20} = \pi/4 \quad (7b)$$

$$\overline{\alpha}_3(\gamma_3+\rho_3+\frac{2\gamma\chi^2\beta_2^2}{(\gamma+\rho)[\gamma+\rho-\overline{\alpha}_3^2\chi^2/(\gamma+\rho)]^2})=\sqrt{2\gamma_3}\beta_3, \qquad (7c)$$

Equation (7) is a five-order equation about $\overline{\alpha}_3$, the numerical solutions could be obtained when the other physical quantities were given. Substitute the numerical solutions $\overline{\alpha}_3$ into Eq. (6), the $\overline{\alpha}_1$ and $\overline{\alpha}_2$ can be given

$$\overline{\alpha}_2=\frac{\sqrt{2\gamma}\beta_2}{\gamma+\rho-\overline{\alpha}_3^2\chi^2/(\gamma+\rho)}, \qquad (8a)$$

$$\overline{\alpha}_1=\frac{\sqrt{2\gamma}\chi\beta_2\overline{\alpha}_3}{(\gamma+\rho)[\gamma+\rho-\overline{\alpha}_3^2\chi^2/(\gamma+\rho)]}, \qquad (8b)$$

Using the boundary condition [35] $\overline{\alpha}_i^{out}=\sqrt{2\gamma_i}\overline{\alpha}_i-\overline{\alpha}_i^{in}$, the conversion efficiency $\eta$ is given by

$$\eta=\frac{\overline{I}_1^{out}}{\overline{I}_3^{in}}=\frac{(\sqrt{2\gamma}\overline{\alpha}_1)^2}{(\beta_3)^2}, \qquad (9)$$

The dynamics of quantum fluctuations can be described by linearizing the equations of motion around the stationary solution by setting

$$\hat{a}_i(t)=\overline{\alpha}_i+\delta\hat{a}_i(t), \hat{a}_i^{in}(t)=\overline{\alpha}_i^{in}+\delta\hat{a}_i^{in}(t), \hat{c}_i^{in}(t)=\delta\hat{c}_i^{in}(t), \qquad (10)$$

Substituting Eq. (10) into Eqs. (5), we obtain the fluctuation dynamics equations

$$\tau\delta\dot{a}_1(t)=-(\gamma+\rho)\delta a_1(t)+\chi\overline{\alpha}_2\delta a_3(t)+\chi\overline{\alpha}_3\delta a_2^+(t)+\sqrt{2\gamma}\delta a_1^{in}(t)+\sqrt{2\rho}c_1^{in}(t), \qquad (11a)$$

$$\tau\delta\dot{a}_2(t)=-(\gamma+\rho)\delta a_2(t)+\chi\overline{\alpha}_1\delta a_3(t)+\chi\overline{\alpha}_3\delta a_1^+(t)+\sqrt{2\gamma}\delta a_2^{in}(t)+\sqrt{2\rho}c_2^{in}(t), \qquad (11b)$$

$$\tau\delta\dot{a}_3(t)=-(\gamma_3+\rho_3)\delta a_3(t)-\chi\overline{\alpha}_1\delta a_2(t)-\chi\overline{\alpha}_2\delta a_1(t)+\sqrt{2\gamma_3}\delta a_3^{in}(t)+\sqrt{2\rho_3}c_3^{in}(t), \qquad (11c)$$

Using the definitions of the amplitude and phase quadratures $X=1/2(a+a^+)$ and $Y=1/2i(a-a^+)$, we obtain the fluctuation of output signal field after Fourier transformations

$$\delta X_1^{out}(\omega)=\frac{1}{C^+}[\eta_{31}^+\delta X_3^{in}(\omega)+\eta_{32}^+\delta X_2^{in}(\omega)+\eta_{33}^+\delta X_1^{in}(\omega) \\ +\kappa_{31}^+\delta X_{c3}^{in}(\omega)+\kappa_{32}^+\delta X_{c2}^{in}(\omega)+\kappa_{33}^+\delta X_{c1}^{in}(\omega)], \qquad (12a)$$

$$\delta Y_1^{out}(\omega) = \frac{1}{C^-}[\eta_{31}^- \delta Y_3^{in}(\omega) + \eta_{32}^- \delta Y_2^{in}(\omega) + \eta_{33}^- \delta Y_1^{in}(\omega) \\ + \kappa_{31}^- \delta Y_{c3}^{in}(\omega) + \kappa_{32}^- \delta Y_{c2}^{in}(\omega) + \kappa_{33}^- \delta Y_{c1}^{in}(\omega)], \quad (12b)$$

where

$$C^\pm = \pm 2\overline{\alpha_1}\overline{\alpha_2}\overline{\alpha_3}\chi^3 + \overline{\alpha_1}^2 \chi^2(\gamma + \rho + i\omega\tau) + \overline{\alpha_2}^2 \chi^2(\gamma + \rho + i\omega\tau) - \overline{\alpha_3}^2 \chi^2(\gamma_3 + \rho_3 + i\omega\tau) \\ + (\gamma + \rho + i\omega\tau)^2(\gamma_3 + \rho_3 + i\omega\tau),$$

$$\eta_{31}^+ = 2\sqrt{\gamma\gamma_3}\chi(\overline{\alpha_2}\gamma + \overline{\alpha_2}\rho + \overline{\alpha_1}\overline{\alpha_3}\chi + i\omega\tau\overline{\alpha_2}),$$

$$\eta_{31}^- = 2\sqrt{\gamma\gamma_3}\chi(\overline{\alpha_2}\gamma + \overline{\alpha_2}\rho - \overline{\alpha_1}\overline{\alpha_3}\chi + i\omega\tau\overline{\alpha_2}) - C^-,$$

$$\eta_{32}^\pm = \pm 2\gamma\chi(\overline{\alpha_3}\gamma_3 + \overline{\alpha_3}\rho_3 \pm \overline{\alpha_1}\overline{\alpha_2}\chi + i\omega\tau\overline{\alpha_3}),$$

$$\eta_{33}^+ = 2\gamma[\gamma\gamma_3 + \gamma_3\rho + \gamma\rho_3 + \rho\rho_3 + \overline{\alpha_1}^2 \chi^2 - \omega^2\tau^2 + i\omega\tau(\gamma + \gamma_3 + \rho + \rho_3)] - C^+,$$

$$\eta_{33}^- = 2\gamma[\gamma\gamma_3 + \gamma_3\rho + \gamma\rho_3 + \rho\rho_3 + \overline{\alpha_1}^2 \chi^2 - \omega^2\tau^2 + i\omega\tau(\gamma + \gamma_3 + \rho + \rho_3)],$$

$$\kappa_{31}^\pm = 2\sqrt{\gamma\rho_3}\chi(\overline{\alpha_2}\gamma + \overline{\alpha_2}\rho \pm \overline{\alpha_1}\overline{\alpha_3}\chi + i\omega\tau\overline{\alpha_2}),$$

$$\kappa_{32}^\pm = \pm 2\sqrt{\gamma\rho}\chi(\overline{\alpha_3}\gamma_3 + \overline{\alpha_3}\rho_3 + \overline{\alpha_1}\overline{\alpha_2}\chi + i\omega\tau\overline{\alpha_3}),$$

$$\kappa_{33}^\pm = 2\sqrt{\gamma\rho}[\gamma\gamma_3 + \rho\gamma_3 + \gamma\rho_3 + \rho\rho_3 + \overline{\alpha_1}^2 \chi^2 - \omega^2\tau^2 + i\omega\tau(\gamma + \gamma_3 + \rho + \rho_3)].$$

Substituting $\langle X_3^{in}\rangle = \langle Y_3^{in}\rangle = \frac{\sqrt{2}}{2}\beta_3$, $\langle X_1^{out}\rangle = \langle Y_1^{out}\rangle = \frac{\sqrt{2}}{2}\overline{\alpha_1^{out}}$ and $\delta^2 X_i^{in} = 1$ into Eq. (3), (21) and (12), we have

$$T_X = \frac{2\gamma\overline{\alpha_1}^2 (C^+)^2}{\beta_3^2 [(\eta_{31}^+)^2 + (\eta_{32}^+)^2 + (\eta_{33}^+)^2 + (\kappa_{31}^+)^2 + (\kappa_{32}^+)^2 + (\kappa_{33}^+)^2]}, \quad (13a)$$

$$T_Y = \frac{2\gamma\overline{\alpha_1}^2 (C^-)^2}{\beta_3^2 [(\eta_{31}^-)^2 + (\eta_{32}^-)^2 + (\eta_{33}^-)^2 + (\kappa_{31}^-)^2 + (\kappa_{32}^-)^2 + (\kappa_{33}^-)^2]}, \quad (13b)$$

### 3. Results and discussion

The dependence of noise figure $T_X, T_Y$ and conversion efficiency $\eta$ on injected idler mode amplitude ($\beta_2$) under different input pump amplitude ($\beta_3$) are shown in Fig. 2. When the input pump amplitude $\beta_3$ is far below the threshold, e.g. $\beta_3 = 0.1\varepsilon_{th}$ in Fig. 2(a), both the noise figure $T_X, T_Y$ and conversion efficiency $\eta$ increase monotonously with the increasing of the injection amplitude $\beta_2$, and they reach the maximum value 1 simultaneously at $\beta_2 = 2.2\varepsilon_{th}$, as shown in Fig. 2(a), which means that almost ideal noise-free quantum state transfer from $a_3^{in}$ to $a_1^{out}$ is realized. If the input amplitude $\beta_2$ is further increased, the $T_X, T_Y$ and $\eta$ will decreased, it shows that for the OPO operating below threshold, an appropriate input amplitude of idler mode (about ten times great than the pump field in this consideration) would be needed for frequency transfer. Note that the noise figure of quadrature amplitude component $T_{X,Y}$ is always equal to conversion efficiency $\eta$ on the condition of Fig. 2(a). When the input pump amplitude is set to be half of the threshold (Fig. 2(b)) or near the threshold (Fig. 2(c)), both $T_X$ and $\eta$ can get to 1 at the same time, but the $T_Y$ drops to below 1. Figure 2 (d) shows that $T_X$, $T_Y$ and $\eta$ will differ greatly with each other, if the OPO is operating above the threshold.

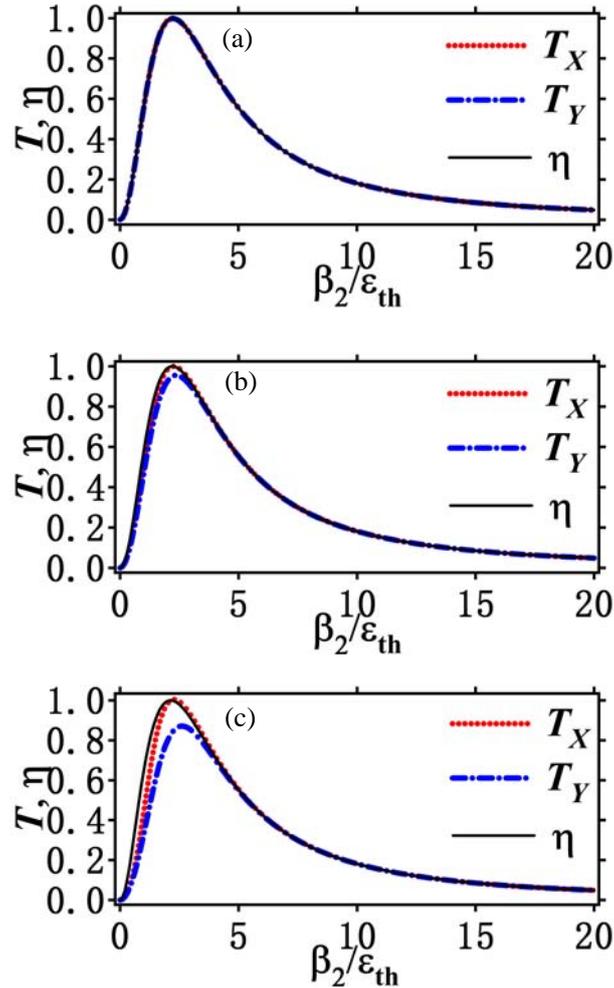

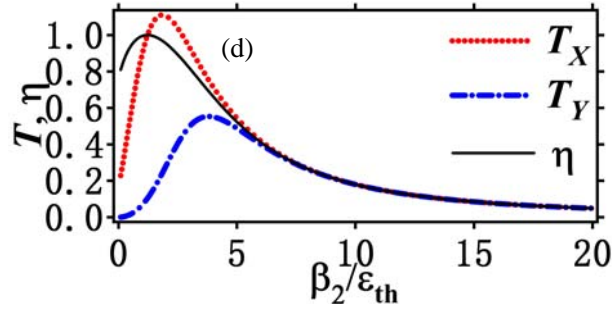

FIG. 2. The noise figure $T_X, T_Y$ and conversion efficiency $\eta$ versus normalized injected idler mode amplitude ($\beta_2 / \varepsilon_{th}$) under different normalized input pump mode amplitude $\beta_3$ (a) $0.1\varepsilon_{th}$ (b) $0.5\varepsilon_{th}$ (c) $0.95\varepsilon_{th}$ (d) $3\varepsilon_{th}$ with $\gamma = \gamma_3 = 0.1$, $\rho = \rho_3 = 0$, $\chi = 0.001$ and $\Omega = \omega\tau/\gamma = 0$.

When the injecting idler amplitude $\beta_2$ is strong enough, to what extent does pump influence the noise figure and conversion efficiency? Figure 3 shows that the conversion efficiency $\eta$ and noise figure $T_X, T_Y$ decrease with increasing the input pump field $\beta_3$. When $\beta_3$ is very weak ($\beta_3 \leq 0.1\varepsilon_{th}$), the three quantities of $\eta, T_X, T_Y$ can catch up to 1 for faithful state transfer, if $\beta_3 > 0.1\varepsilon_{th}$, though the $T_X$ and $\eta$ remain approximately constant during this period, the noise figure of quadrature phase component $T_Y$ will decrease rapidly.

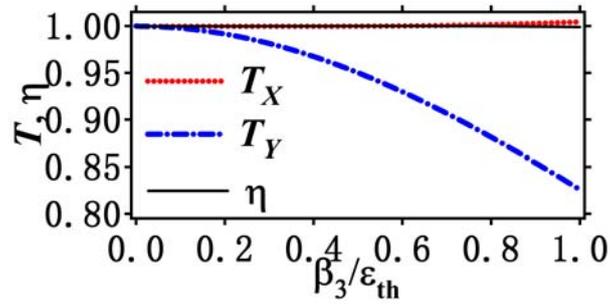

FIG. 3. The noise figure $T_X, T_Y$ and conversion efficiency $\eta$ versus $\beta_3$ with $\beta_2 = 2.2\varepsilon_{th}$, $\gamma = \gamma_3 = 0.1$, $\chi = 0.001$ and $\Omega = 0$.

Figure 4 shows the noise figure of quadrature components $T_X$ and $T_Y$ versus the normalized analyzing frequency $\Omega$ when the OPO is below threshold for $\beta_3 = 0.1\varepsilon_{th}$. It can be found that $T_{X,Y}$ are always frequency-independent.

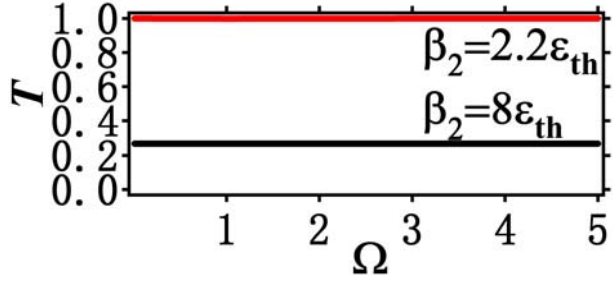

FIG. 4. The noise figure $T_X, T_Y$ versus normalized frequency $\Omega$ with $\beta_3 = 0.1\varepsilon_{th}$, $\gamma = \gamma_3 = 0.1$, $\rho = \rho_3 = 0$ and $\chi = 0.001$.

From the above analysis, it is evident that the perfect quantum state transfer with down-conversion in OPA can be obtained as long as the OPA is operated far below threshold, and an idler injection with proper amplitude. Note that in this discussion, we set the parameter of extra intracavity loss $\rho_i = 0$. If this unavoidable loss is included, the noise figure for quadrature components and conversion efficiency will decrease (see Fig. 5). In order to get high performance quantum state transfer, we have to improve the OPA that the extra loss should less than the transmission loss from the coupling mirror, $\rho_i \ll \gamma$ (we set $\gamma = \gamma_3 = 0.1$ in Fig.5).

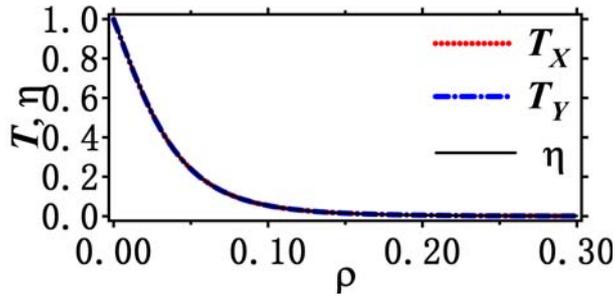

FIG. 5. The noise figure $T_X, T_Y$ and conversion efficiency $\eta$ versus extra cavity loss $\rho = \rho_3$ with $\beta_3 = 0.1\varepsilon_{th}$, $\beta_2 = 2.2\varepsilon_{th}$, $\gamma = \gamma_3 = 0.1$, $\chi = 0.001$ and $\Omega = 0$.

**4. Conclusion**

In conclusion, the frequency transfer efficiency and the noise property of the quantum state of the quantum frequency down-conversion generated in an NOPA has been theoretically discussed, which shows that the well behaved quantum state transfer can be obtained when the NOPA is operated far below threshold with the injection of a strong idler mode. By successfully using the NOPA technology, the experimental realization of a high efficiency quantum frequency transfer may be achieved. This scheme for quantum state transfer between different frequencies may have potential application in quantum network and quantum computation processes.

Acknowledgements: This work is supported in part by the NSFC (No. 10974126, 60821004), National Basic Research Program of China (No. 2010CB923102), Research Project Supported by Shanxi Scholarship Council of China.